# Integrated recurrent optical spectral slicer for equalization of 100-km C-band IM/DD transmission

I. Teofilovic[(1)], K. Sozos[(2)], H. Liu[(3)], S. Malhouitre[(4)], S. Garcia[(4)], G. Sarantoglou[(5)],
P. Bienstman[(6)], B. Charbonnier[(3)], C. Mesaritakis[(5)], C. Vigliar[(1)],
P. Petropoulos[(3)], A. Bogris[(2)], F. Da Ros[(1)]

[(1)] Department of Electrical and Photonics Engineering, Technical University of Denmark, Ørsteds Plads, DK-2800 Kgs. Lyngby, Denmark (isteo@dtu.dk, fdro@dtu.dk)
[(2)] Department of Informatics and Computer Engineering, University of West Attica, Aghiou Spiridonos, 12243, Egaleo, Athens, Greece (ksozos@uniwa.gr, abogris@uniwa.gr)
[(3)] Optoelectronics Research Centre, University of Southampton, Southampton SO17 1BJ, United Kingdom, (hl19n23@soton.ac.uk, pp@orc.soton.ac.uk)
[(4)] CEA-Leti, University of Grenoble Alpes, Grenoble, France (benoit.charbonnier@cea.fr)
[(5)] Department of Biomedical Engineering, University of West Attica, Aghiou Spiridonos, 12243, Egaleo, Athens, Greece, (cmesar@uniwa.gr)
[(6)] Ghent-University/imec, Technologiepark-Zwijnaarde 126, B-9052 Ghent, Belgium (peter.bienstman@ugent.be)

**Abstract** A silicon-photonics recurrent spectral filter is designed, fabricated, and system tested to provide optical pre-processing in a 32-GBd PAM-4 C-band transmission. Performance below FEC is reported for up to 100 km reach. ©2025 The Author(s)

## Introduction

Short-reach (≤ 100 km) optical communications are driven by the need for low-cost and low energy consumption solutions. Currently, such requirements are still easier to meet with intensity-modulation direct-detection (IM/DD) systems. However, the drive for increasing the symbol rate/lambda clashes with the power fading limitations introduced by the interplay of dispersion effects and square law detection especially away from the O-band and ultimately limits the transmission reach [1-3].

The use of nonlinear equalizers, including complex machine-learning driven methods [3,4], can be effective in recovering part of the lost reach by mitigating intersymbol interference. Nevertheless, the loss of the phase information during DD results in inferior performance compared to coherent systems.

As alternatives to purely digital equalizers, solutions incorporating pre-processing in the optical domain have been proposed [5-9].

Among them, the recurrent optical spectral slicing (ROSS) technique relies on applying multiple frequency-shifted infinite impulse response filters (nodes) to slice the signal spectrum. Each slice is independently detected, and the signal is reconstructed and equalized in the digital domain with a simple linear feed-forward equalizer (FFE) [7]. The frequency diversity achieved by using non-frequency-overlapping ROSS nodes allows for mitigating the impact of power fading. Substantial performance improvement has been shown both in numerical investigations [7], and in a preliminary experimental validation using a programmable photonic integrated circuit (PIC) [10] implementing the ROSS node [8].

However, the use of a general-purpose programmable PIC introduced increased insertion losses and limited the freedom in the filter design due to the need to comply with the granularity provided by the programmable mesh [8, 10].

In this work, we design, fabricate, and test two ad-hoc tuneable PIC filters tailored for a ROSS receiver. The proposed filters are based on an asymmetric Mach-Zehnder interferometer architecture incorporating delayed feedback from the output to the input. The devices used in the experiment consisted of two variants in terms of time difference (ΔT) values between the MZI's arms and they are evaluated in terms of providing equalization for a 32-GBd PAM-4 signal. A transmission reach of up to 100 km is shown for a 2-node ROSS receiver based on the short ΔT design. Bit error ratio (BER) performance well below the KP4 hard-decision forward error correction (HD-FEC) threshold (BER ≤$2.25×10^{-4}$ [11]), at 50 km and below the 400ZR C-FEC threshold (BER≤$1.25×10^{-2}$, [12]) at 100 km is experimentally validated.

## PIC Design and Fabrication

The ROSS node design is based on an asymmetrical MZI with the feedback loop between output and input corresponding to a delay of approx. 19 ps. The design choice was based on our previous theoretical analysis [7] and aimed at improving on the limitations introduced by the programmable photonic platform used in [8]: (1) the limited flexibility in free-spectral range (FSR) since delays need to be multiples of the programmable unit cell = 11.25 ps [10], (2) the loss in the feedback path (> 1dB) which directly limits the feedback strength [8], and (3) high in- and out-coupling loss introduced by the PIC design [10]. For our proposed filters, two values of ΔT are considered: approx. 44 ps and 23 ps. In order to tune the filter position with respect to the optical channel under test, a thermo-optic phase-shifting element is placed in the feedback arm. The tunability of the phase shifter allows realizing different ROSS nodes, i.e., selecting different spectral slices of the signal depending on the phase shift, with the same device design. A sketch of the two designs is provided in Fig. 1, including the vertical grating couplers used for in- and out-coupling.

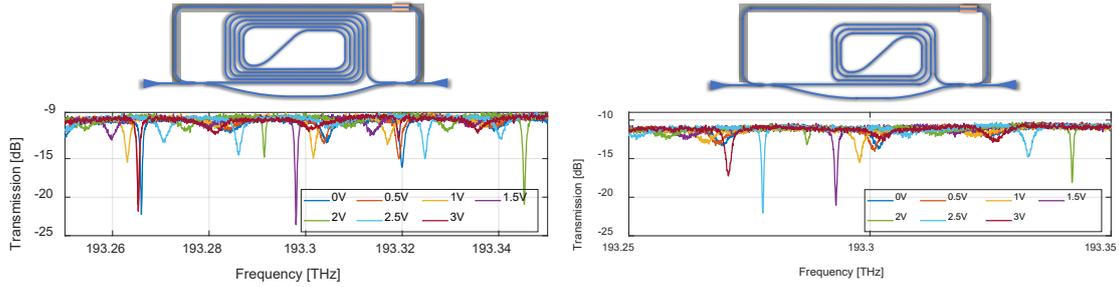

**Fig. 1:** Design and measured transmission of the two filters for different applied voltages: (a) long, and (b) short ΔT.

The designed silicon chip was fabricated on CEA-Leti's 300mm silicon Photonics platform using immersion lithography. Only passive silicon layers were fabricated (waveguides, grating couplers and splitter/couplers), and they were complemented by a micro-heater layer in TiTiN and a single layer of copper for electrical interconnections using Tungsten vias.

### PIC characterization

The coupling loss of the fabricated PIC was estimated to 2.5 ± 0.2 dB/coupler and the waveguide propagation loss to 0.32 ± 0.1 dB/cm. Manual polarization alignment was performed for all the measurements presented due to the polarization sensitivity of the vertical grating couplers.

The transfer function of the two filter designs was characterized using a 5-nm polarized amplified spontaneous emission (ASE) source and feeding the PIC output to a high-resolution (180 MHz) optical spectrum analyzer (OSA). Cleaved fibers have been used for in- and out-coupling through the vertical grating couplers. Voltage to the heater was applied through needle probes to vary the phase of the feedback loop. The transfer functions are shown in Fig. 1 for different values of voltage applied. The curves highlight the sharp features, and high extinction ratio (up to 15 dB). By tuning the applied voltage, the filter shapes are not only frequency shifted, but a rich set of spectral shapes can also be achieved. This is key in providing the spectral diversity needed for the ROSS technique and is enabled by this PIC, especially due to the decreased losses in the feedback loop (estimated < 0.1 dB) compared to an equivalent structure configured in a programmable PIC (> 1 dB in the implementation of [8]). The overall insertion losses are 9.8 and 10.4 dB for long and short ΔT, respectively.

### System performance

The experimental setup to evaluate the achievable performance with the proposed filter design is shown in Fig. 2. A 32-GBd PAM-4 electrical signal was generated by an arbitrary waveform generator (AWG) operating at 64 GSa/s, amplified by linear electrical drivers and used to drive a Mach-Zehnder modulator (MZM) encoding the information onto an optical carrier from an external cavity laser (ECL, λ= 1550.1 nm). The optical signal was input into a spool of single-mode fiber (SMF) with the length varied between 50 and 100 km, and at a launch power of 0.5 dBm. At the output of the SMF, the signal was amplified by an erbium-doped fiber amplifier (EDFA) to 11 dBm before inputting it into the PIC-based ROSS node with the same cleaved fiber setup used for the the transfer function measurements. At the output of the PIC, a second EDFA followed by a 200-GHz band pass filter (BPF) and a photodetector without a trans-impedance amplifier emulated a pre-amplified receiver. A digital storage oscilloscope (DSO) was used to acquire the time-domain

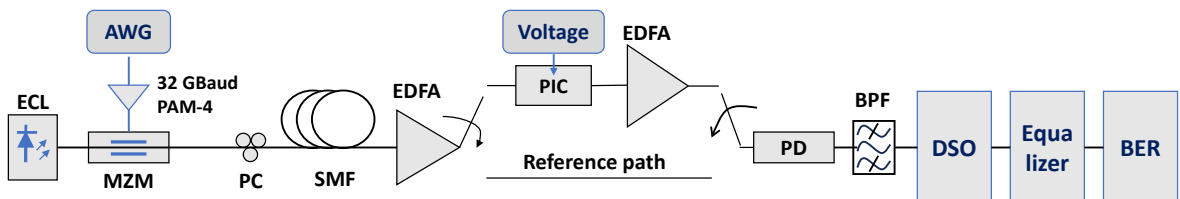

**Fig. 2:** System setup for evaluating the two filter designs for a ROSS receiver.

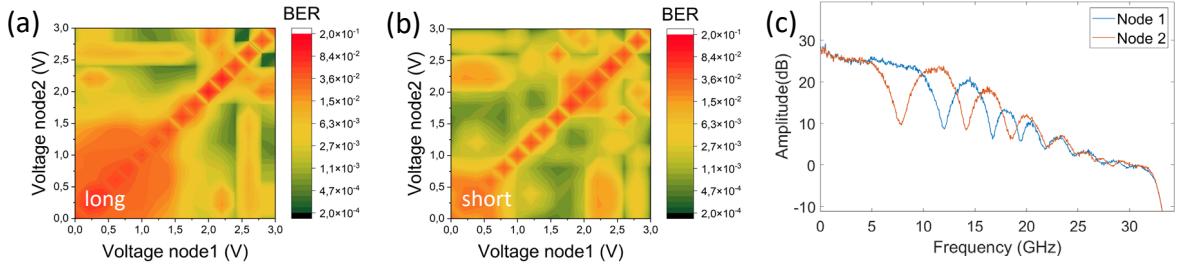

**Fig. 3:** BER performance for 50km transmission: BER as a function of the filter pairs for a long (a) and short (b) ΔT; Electrical spectra from node 1 and node 2 considering the best performing filter pair for the short ΔT design.

traces for offline processing, consisting of equalization and BER counting. Due to the lack of a multi-node PIC, the two ROSS nodes were emulated by sequentially measuring DSO traces for different voltages applied to the PIC, i.e., different filter transfer functions. The 2-node ROSS receiver was then constructed by synchronizing the traces from two nodes and feeding them simultaneously through a 62 T-spaced-taps (31 symbols/filter) feedforward equalizer (FFE). Different node pairs were analyzed, and the performance achievable at 50 km transmission for the long and short ΔT is reported in Fig. 3 (a) and (b), respectively. The heatmaps show the BER as a function of the voltage applied to the heater for the two nodes. Consistent with expectations and previous demonstrations [7,8], frequency overlapping filters (anti-diagonal values) provide poor BER performance. In contrast, once the two filters provide nonoverlapping spectral slices, substantial BER improvement can be seen for both filter designs. An example of received electrical spectra from the two nodes leading to the best BER are shown in Fig. 3(c) for the short ΔT design. The

BER performance measured with the pre-amplified receiver directly at the SMF output, i.e., bypassing an EDFA and the PIC (reference path in Fig. 2), is also shown for two equalizers: a FFE equalizer and a nonlinear neural-network-based equalizer. The number of taps was kept identical to the ROSS receiver. Finally, the 50-km BER results for the previously considered programmable PIC ([8]) are also reported. For this realization, the filter was implemented as an asymmetrical MZI with feedback, but the path difference ΔT was set to 45 ps, the feedback length to 22.5 ps, and the resulting insertion loss was above 20 dB.

As can be seen, poor BER performance is obtained even at 50 km when no optical pre-processing is applied, even when a nonlinear equalizer is used (BER~$2×10^{-2}$). Using the two-node ROSS receiver based on the programmable PIC, improves the performance by almost an order of magnitude (BER~$3×10^{-3}$). However, the additional loss in the feedback path (>1 dB vs. < 0.1 dB), the overall higher insertion loss (~10 dB), and the constrained FSR design do not provide as much improvement as the fabricated PICs reported in this work. Both designs (long and short) allow for an additional order of magnitude improvement (BER~$2×10^{-4}$ < KP4 FEC threshold at 50 km). The improvement is consistent over the transmission distances, up to the 100 km tested, with slightly better performance achievable with the short ΔT design (in line with our previous analysis using a programmable PIC and varying the feedback length [8]). Considering such a filter, the BER can be kept below the 400ZR C-FEC threshold even for 100 km transmission.

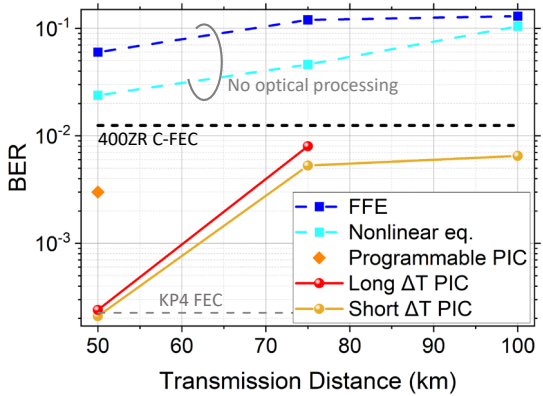

**Fig. 4:** BER performance as a function of the transmission distance: no-optical processing (FFE and nonlinear equalizers); ROSS with programmable PIC [8], ROSS with long and short ΔT.

spectra confirm that the performance improvement originates from the spectral selectivity.

The performance achievable with the best node pairs as a function of the transmission distance is shown in Fig. 4 for the long and short ΔT. Note that the best node positions are consistent over transmission distance. As benchmarks, the

## Conclusions

We have designed, fabricated, and characterized the performance of a recurrent spectral slicing filter implementing a two-node ROSS receiver. The proposed designs, combined with a simple digital FFE, allow for up to two orders of magnitude improvement in BER compared even with complex nonlinear equalizers. BER performance below the KP4 FEC and 400ZR C-FEC thresholds is achieved up to 50 and 100 km of SMF transmission, respectively.

## Acknowledgements

This work has been supported by Horizon Europe projects PROMETHEUS under grant agreement 101070195, the H2020 project NEoteRIC under grant agreement 871330, the Villum Foundations project OPTIC-AI under grant agreement VIL29344, and the UK's EPSRC HASC Telecommunications Hub (EP/X040569/1).